\documentclass[aps,pre,twocolumn]{revtex4-1}

\usepackage{amsfonts}
\usepackage{amsmath}
\usepackage{amssymb}
\usepackage{bbm}
\usepackage{bbold}
\usepackage{dcolumn}
\usepackage{epsfig}
\usepackage{gensymb}
\usepackage{graphics}
\usepackage{graphicx}
\usepackage{latexsym}
\usepackage{textcomp}

\usepackage{color}
\usepackage[dvipsnames]{xcolor}

\usepackage{textgreek}
\usepackage{MnSymbol}

\newcommand{\al}{\alpha}
\newcommand{\alb}{\br{\al}}

\newcommand{\be}{\beta}
\newcommand{\beb}{\br{\be}} 
\newcommand{\bee}{\be^{\rm{e}}}

\newcommand{\br} [1]{\bar{#1}}

\newcommand{\crl}{correlation}

\newcommand{\dif}  {D}
\newcommand{\difr} {\mathcal{D}}
\newcommand{\difs} {\sttar{\dif}}

\newcommand{\dm}{d}

\newcommand{\dr}{D_\mathrm{r}}
\newcommand{\drm}[1]{D_{\mathrm{r},m}}
\newcommand{\eqdef}  {\equiv}

\newcommand{\fac}  {K}    

\newcommand{\lc}{l_\mathrm{c}}

\newcommand{\Lav}{\langle L \rangle}
\newcommand{\liou}{\mathcal{L}}

\newcommand{\por}{\varphi}
\newcommand{\porb}{\br{\por}}

\newcommand{\rat}{\mathcal{R}}

\newcommand{\sfs}{\mathcal{S}}

\newcommand{\sttar} [1]{#1^*}
\newcommand{\ssw}{\mathrm{s}}
\newcommand{\suml}{\sum_{l=\pm1}}   
\newcommand{\SM}{\cite{sm}}   
\newcommand{\taup}{\tau_\mathrm{p}}
\newcommand{\taups}{\sttar{\taup}}
\newcommand{\taur}{\tau}
\newcommand{\taurs}{\sttar{\tau}}
\newcommand{\tauw}{\tau_\ssw}

\newcommand{\thetat}{\theta_{\rm{t}}}

\newcommand{\T}{\mathcal{T}}      
\newcommand{\vo}{v_\mathrm{o}}

\newcommand{\Tunit}{t_\mathrm{u}}

\definecolor{darkred}{rgb}{0.6, 0.0, 0.0}

\begin{document}

\title{Universal law for the dispersal of motile microorganisms in porous media}

\author{T.~Pietrangeli$^{1,*}$}
\author{R.~Foffi$^{2,*}$}
\author{R.~Stocker$^2$}
\author{C.~Ybert$^1$}
\author{C.~Cottin-Bizonne$^1$}
\author{F.~Detcheverry$^1$}
\affiliation{$^1$University of Lyon, Universit\'{e} Claude Bernard Lyon 1, CNRS, Institut Lumi\`{e}re Mati\`{e}re, F-69622, Villeurbanne, France}
\affiliation{$^2$Department of Civil, Environmental and Geomatic Engineering, Institute of Environmental Engineering, ETH Zurich, Zurich, Switzerland}

\begin{abstract}
%
Dispersal is essential to the plethora of motile microorganisms living in porous environments, 
yet how it relates to movement patterns and pore space structure remains largely unknown. 
Here we investigate numerically the long-time dispersal of a run-and-tumble microorganism 
that remains trapped at solid surfaces and escapes from them by tumbling. 
We find that dispersal 
and mean run time are connected by a universal relation,
that applies for a variety of porous microstructures and swimming strategies. 
We explain how this generic dependence   originates in 
the invariance of the mean free path with respect to the movement pattern, 
and we discuss the optimal strategy that maximizes dispersal.    
Finally, we  extend our approach to microorganisms moving along the surface.  
Our results 
provide a general framework to quantify dispersal 
that works across the vast diversity of movement patterns and porous media. 
\end{abstract}

\date{\today}

\maketitle


%
Among the $10^{30}$ prokaryotes that populate the Earth,  a majority live in  oceanic and terrestrial subsurfaces,  made of sediments,  soil or rocks~\cite{Whitman_pnas-1998,Kallmeyer_pnas-2012}. 
With myriads of bacteria also inhabiting higher organisms~\cite{Sender_plosb-2016}, 
porous media is a widespread habitat for microbial life~\cite{Jin_bpr-2024,Shrestha_csb-2023}. 
The strategies microorganisms adopt to navigate their environment impact their ability 
to access resources~\cite{Frankel_elife-2014}, invade  
new areas~\cite{Finkelshtein_mbio-2015,Kurkjian_ploscb-2021}, 
and ultimately contribute to determine their survival~\cite{Coyte_pnas-2016}. 
Therefore, establishing how navigation strategies regulate dispersal 
in relation to the physical environment 
is  key to understand the ecological success of many species of bacteria.
Additionally, 
predicting bacterial dispersal is important because it plays a 
crucial role in infections~\cite{Zegadlo_ijms-2023}, food contamination by pathogens~\cite{Ribet_microbesinf-2015}, targeted drug-delivery in tumors~\cite{Gupta_vaccines-2021}, rhizosphere enhancement for plant growth~\cite{Compant_sbb-2010} 
and the bioremediation of porous aquifers~\cite{Ginn_awr-2002}.

Understanding the dispersal of motile microorganisms poses at least two challenges.
The first  is the existence of a vast parameter space. 
Bacteria generally swim in random walks 
where nearly straight runs are punctuated by reorientation events, 
which  exist in several types and define a repertoire of swimming 
strategies~\cite{GrognotTaute_comb-2021,Stocker_pnas-2011,Jarrell_natrevmb-2008,Lauga_rpp-2009,Herzog_aem-2012}. 
The diversity of porous microstructures, from rocks to tissues and body gels, is no less daunting, with pores widely varying in morphology and spanning micrometers to millimeters in size~\cite{Jin_bpr-2024}.
In spite of a recent surge of research 
on both non-tumbling~\cite{Morin_pre-2017,BrunCosmeBruny_jcp-2019,Jakuszeit_pre-2019,Dehkharghani_comphys-2023}
and tumbling microorganisms
\cite{Duffy_bpj-1995,Licata_bpj-2016,Bhattacharjee_natcom-2019,Kurzthaler_natcom-2021,Irani_prl-2022,Rizkallah_prl-2022,Lohrmann_pre-2023,Saintillan_pre-2023,Pietrangeli_prr-2024,Jakuszeit_pre-2024}, 
most of the parameter space remains unexplored. 
The second challenge is that little generic finding has emerged. 
The one exception is the existence, across different systems, of an optimal persistence time 
at which dispersal is maximal~\cite{Licata_bpj-2016,Bertrand_prl-2018,Kurzthaler_natcom-2021,Irani_prl-2022,Rizkallah_prl-2022,Lohrmann_pre-2023,Pietrangeli_prr-2024},
yet there is no overarching principle to predict this maximum. 
In this context, 
an intriguing proposition put forward in a 
preprint by Mattingly~\cite{Mattingly_arxiv-2023}   
is that the microstructure can be ``largely forgotten'', 
in the sense that only a small set of  features is relevant. 
How general this finding is, however, remains unknown, because it was reached for a specific microstructure and swimming strategy.
Overall, 
it is unclear how knowledge gained 
in idealized porous media mostly used to date  
-- typically arrangements of spheres or disks -- ,  
can be transferred to the great diversity of microstructures found in the natural world. 

\begin{figure}[b!] 
\center
\includegraphics[width=8cm]{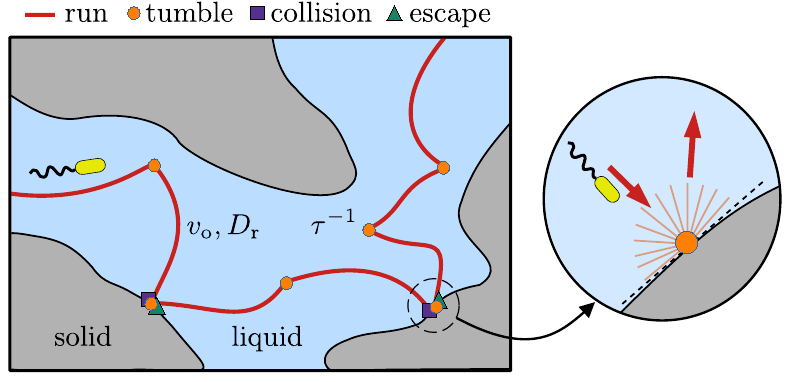} 

\vspace*{-2mm}

\caption{
{\bf Model of run-and-tumble microorganism in a porous medium} 
(see text for parameter definition). 
After a collision with the surface, the microorganism escapes by tumbling, 
with a direction randomly sampled in the available half-space 
(orange rays in the inset). 
}  
\label{fig:1}
\end{figure}

%

\begin{figure*}[t!]
\includegraphics[width=0.999\textwidth]{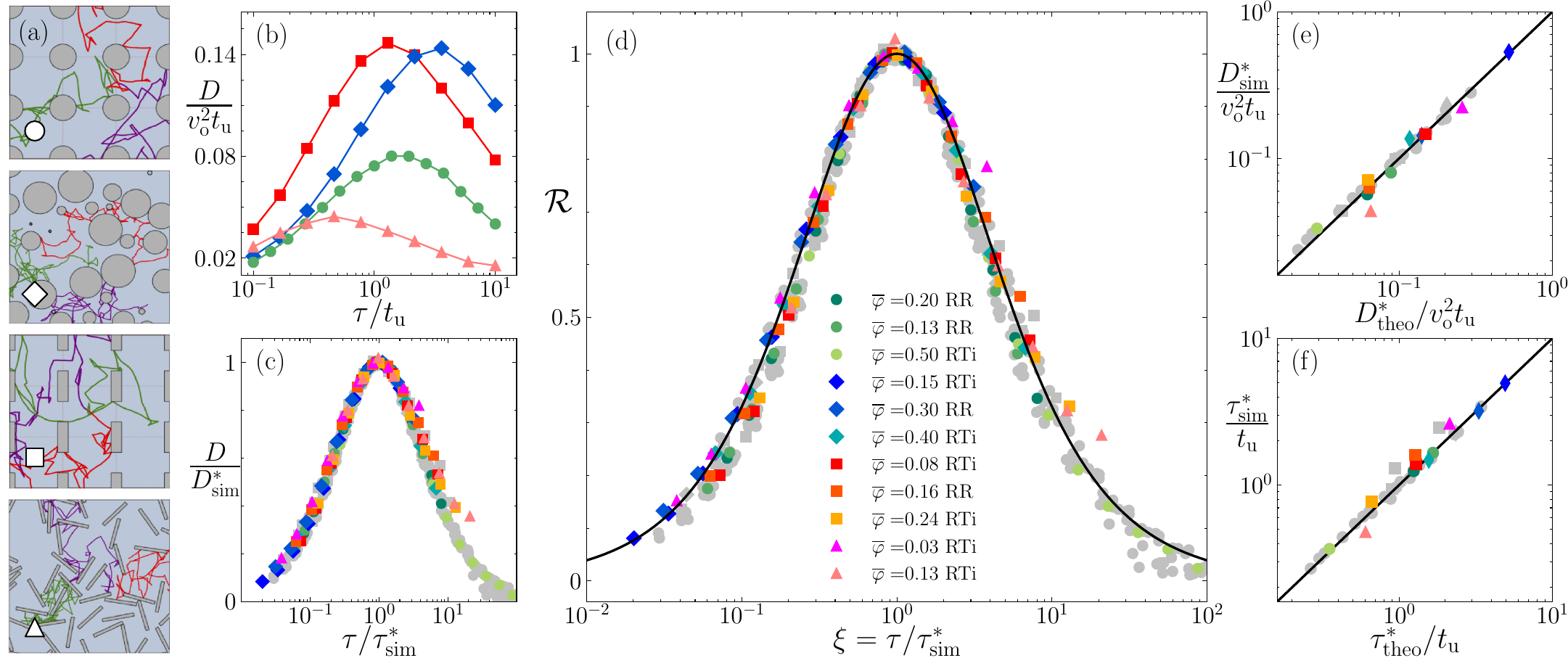} 
\caption{
{\bf Diffusivity of a run-and-tumble microorganism in porous media.}
(a)~Simulated motion within different microstructures: 
ordered disks~($\circ$), disordered polydisperse disks~($\smalldiamond$), ordered rectangles~($\smallsquare$) and disordered rods~($\smalltriangleup$), 
with solid fractions $\porb=1-\por=0.25$, $0.3$, $0.24$ and $0.13$ respectively. 
(b)~Diffusivity as a function of the mean run time for four systems.  
The timescale $\Tunit$ depends on the microstructure \SM.
(c)~Diffusivity curves %
rescaled by the value and time of their maximum. 
Grey points show additional data, detailed in Table~I~\SM, including different microstructures, swimming strategies, rotational diffusivities, solid fractions between 0.01 and 0.5, 
and a model  with partial trapping.  
(d)~The modified diffusivity ratio $\rat$
for all investigated models
collapses on a universal curve (black line), 
defined by Eq.~\eqref{eq:rat}. 
The legend applies across panels (b)-(f).  
Here $\dr=0.1\,\Tunit^{-1}$. 
(e-f)~Comparison between simulations ($y$-axis) and theory ($x$-axis) for 
(e) the maximum diffusivity~$\difs$ and 
(f) the optimal mean run time~$\taurs$.
}  
\label{fig:2}
\end{figure*}

In this work, we show that the dispersal of motile microorganisms in porous media 
has a universal character. 
We study a run-and-tumble microorganism, 
which stops moving upon encountering a solid surface but can escape from it by tumbling. 
 We find numerically that, across diverse microstructures and swimming strategies, 
 the dispersal is related to the mean run time by a universal dependence, that we derive in an analytical model.  
We explain why 
one can forget almost everything about the porous medium when predicting the microorganism diffusivity.     
The  key insight 
is the invariance of the mean free path~\cite{Bardsley_siamjam-1981,Blanco_epl-2003}. 
As a result,   
what matters for dispersal is only the ratio between 
the accessible volume and the amount of interface. 
Besides, we show that the dispersal law is also relevant for microorganisms moving along surfaces.
Our results show that, despite the diversity of porous structures and motion patterns in natural systems,             
dispersal can be understood within a common and remarkably simple framework. 

%
We consider a  run-and-tumble (RT) microorganism moving  in  a porous medium~(Fig.~\ref{fig:1}). 
It has velocity~$\vo$, is subject to rotational diffusion with coefficient~$\dr$ and tumbles with rate $\taur^{-1}$, with a distribution of reorientation angles that is nonchiral but otherwise arbitrary.  
The model includes as particular cases 
 two  swimming strategies: 
 run-reverse (RR), where reorientation events involve perfect reversal of swimming direction, 
 and  RT with isotropic reorientation (RTi). 
We assume 
the simplest behavior at the surface~\cite{Junot_prl-2022}:   
the microorganism has vanishing velocity there but at each tumble, 
it can escape with a probability~$\eta^{-1}$ 
and a direction uniformly distributed in the free half-space (Fig.~\ref{fig:1} inset). 
Dispersal 
is characterized by the diffusivity  $\dif = \lim_{t \to \infty} M(t)/2 \dm t$, 
with $M(t)$ the mean-square displacement at time~$t$ and~$\dm$ the space dimension.  
We performed agent-based simulations~\cite{Foffi_zenodo-2025} 
to determine the 
diffusivity as a function of the mean run time~$\taur$ in different porous environments, both ordered and disordered~(Fig.~\ref{fig:2}a). 
We found that, for all environments and swimming strategies tested, 
there is an optimal run time $\taurs$ for which the diffusivity $\dif$ reaches a maximum (Fig.~\ref{fig:2}b).
Moreover, when rescaled by the diffusivity maximum~$\difs$ and by the optimal mean run time~$\taurs$, 
all the data collapse onto a master curve.
We now explain this universal behavior.


\paragraph*{Minimal model of diffusivity and Cauchy universality.}
We propose a simple  model of microbial dispersal in a porous medium, 
whose main approximation is to discard the complex correlations between microorganism trajectory 
and  microstructure. 
Specifically, we assume that encounters with solid surfaces 
occur  along the trajectory as a Poissonian process with rate~$\T^{-1}$ 
and can therefore be treated as a second type of tumble~\cite{Jakuszeit_pre-2024}.  
We then derive the diffusivity of the microorganism as  \SM 
\begin{align}
\dif  
      =  \frac{\T}{\T+\tauw}  \;   \frac{\vo^2/d}{\dr' + \alb \taur^{-1} +  \beb \T^{-1} }  
         \, \fac(\porb).             
         \label{eq:model}   
\end{align}
Here, 
$\tauw = \eta \taur$ is the mean time spent at the surface after an encounter,    
$\dr' \eqdef (d-1) \dr$, 
$\bar{\alpha} \eqdef 1-\alpha$, 
$\bar{\beta} \eqdef 1-\beta$  and 
$\bar{\varphi} \eqdef 1-\varphi$. 
$\al \eqdef \langle \cos \thetat \rangle$ 
is the mean cosine of the reorientation angle~$\thetat$  --  henceforth called ``correlation" for short --   
induced by a tumble~\footnote{$\al=0$ for RTi and $-1$ for RR.} 
and $\be$  is similarly the ``correlation" of reorientation  induced by an encounter with the surface.    
$\fac(\porb)$ is a correction factor that depends on the porosity~$\por$ and the microstructure. 
Each of the three factors in Eq.~\eqref{eq:model} encapsulates one physical effect governing dispersal. 
The first factor is the fraction~$\nu$ of time spent moving. 
The second factor captures the three independent processes 
-- rotational diffusion, tumbles and surface encounters --  responsible for orientational decorrelation, whose rates are additive. 
Finally, the third factor accounts for the excluded volume and correlation of the microstructure.   
$\fac(\porb)$ is chosen so that the diffusivity is correct in the limit of 
Brownian motion~\footnote{
Brownian motion is recovered in the limit $\taur \to 0$ and $\vo \to \infty$ while keeping $\vo^2 \taur$  finite.
} ; 
$\fac(\porb)$  is known in several microstructure~\cite{Mangeat_jcp-2020,Novak_jcp-2011,book_Torquato-RandomHeteroMat} 
and, at low solid fraction~$\porb$, 
$\fac(\porb) = 1-\porb/(d-1)+ \mathcal{O}(\porb^2)$ holds for arbitrary material~\SM. 
To complete the model, one then needs only to specify the mean time~$\T$  
between two  surface encounters. 

%
The surface encounter time~$\T$ has a surprisingly simple expression.  
Since velocity is constant in modulus,  
$\vo \T$ is the mean free path $\Lav$,  
defined as the trajectory length between two successive contacts with the surface. 
The mean free path possesses an invariance property, also called Cauchy universality~\cite{Artuso_arxiv-2024}, 
which states that it is equal to the mean chord length $\lc$ of the medium, giving
\begin{align}
 \vo \T =  \Lav = \lc = \sigma_\dm \,\frac{\Omega}{\partial \Omega} = \sigma_\dm \, \frac{\por}{\sfs}, 
\label{eq:lc}   
\end{align}
with $\sigma_d=\pi$ and $4$ for $\dm=2$ and $3$, respectively,   
$\Omega$ the volume of porous space,  
$\partial \Omega$ the amount of solid interface 
and $\sfs$ the specific surface of the material. 
First written for ballistic motion~\cite{book_Santalo-IntroIntegralGeom,book_Torquato-RandomHeteroMat,Chernov_jsp-1997}, 
Eq.~\eqref{eq:lc}  
actually holds true in more general conditions~\cite{Artuso_arxiv-2024,Shukla_pre-2019},  
that are still under investigation \cite{Zamora_jcp-2024}, 
but 
include random motion with reorientation events 
that may be  anisotropic, inhomogeneous 
or even non-Poissonian~\cite{Bardsley_siamjam-1981,Blanco_epl-2003,Mazzolo_jmp-2014}.  
Cauchy universality is thus applicable to our generic run-and-tumble motion.



%
\paragraph*{Diffusivity  master curve and maximum.}
Knowing that the encounter time~$\T$ is a purely geometric quantity 
independent of the mean run time $\taur$, 
Eq.~\eqref{eq:model} gives
\begin{align}
\frac{\dif}{\difs} &= 
 \frac{(2+c)\xi}{1 + c \xi + \xi^2}, \quad \xi \eqdef \frac{\taur}{\taurs}, 
\label{eq:dif}
\end{align}
where $\xi$ is a rescaled mean run time. 
The diffusivity reaches a maximum $\difs$ at the mean run time $\taurs$, with
\begin{equation}
\difs = \sqrt{a b} \fac(\porb)/(2 + c)\alb \dm, \quad \taurs = \sqrt{a b}, 
\label{eq:max}
\end{equation}
and $a \eqdef \T/\eta$, $b \eqdef \alb/(\dr + \beb/\T)$ and  $c \eqdef (a+b)/\sqrt{a b}$. 
The dependence of $\dif/\difs$  on $c$  
can be entirely accounted for by introducing 
the modified diffusivity ratio
\begin{align}
\rat \eqdef 4 \left[ 2 -c + (2+c)   \frac{\difs}{\dif} \right]^{-1} 
= \frac{4\xi}{(1 +  \xi)^2 }, 
\label{eq:rat}
\end{align}
which is a function of $\xi$ only. 
For the 38 parameter combinations tested, 
that differ in swimming strategy, rotational diffusion, porosity or morphology of the medium, 
the diffusivity values from simulations collapse,  
without any free parameter, 
on the master curve $\rat(\xi)$ from Eq.~\eqref{eq:rat} (Fig.~\ref{fig:2}d)~\footnote{
Equation~(3) includes a dependence on the $c$~parameter
but because $c$ varies over a limited range of values across different simulations,  
an approximate collapse is seen for $\dif/\difs$ in Fig.~\ref{fig:2}c.}. 
Additionally, 
the simulation results for~$\difs$ and~$\taurs$ 
also align closely with the theoretical predictions from Eq.~\eqref{eq:max} (Fig.~\ref{fig:2}e-f).   


%
The diffusivity maximum occurs because diffusivity increases as  $\dif \sim \taur$ for short runs   
and decreases as $\dif \sim  (\eta \taur)^{-1}$ for long runs, 
since then most time is spent at the surface waiting to escape. 
The optimal mean run time exhibits two regimes. 
When  rotational diffusion is strong ($\dr \T \gg 1 $), 
 $\taurs \simeq \sqrt{\alb T /\dr\eta}$ 
is the harmonic mean between the surface encounter time and the rotational diffusion time, 
a trade-off between the two processes driving the orientation decorrelation. 
When  rotational diffusion is  negligible ($\dr \T \ll 1 $), 
$\taurs \simeq \T \sqrt{\alb/\eta \beb}$ is controlled by the surface encounter time.
Here, the optimal mean run length $\vo \taurs$ is dictated   
by the mean free path~$\lc=\vo\T$, since it realizes the best compromise 
between efficient transport in porous space 
and the penalty of being blocked at the surface.

\begin{figure*}[t!]
\includegraphics[height=4.3cm  ]{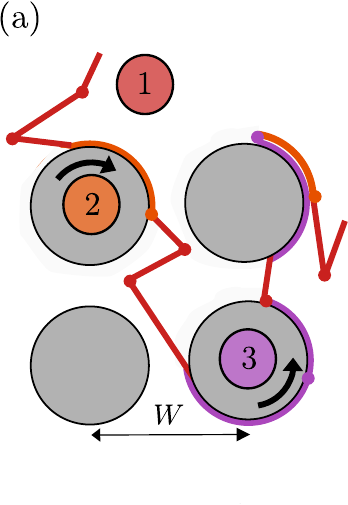} \hspace*{1mm}
\includegraphics[height=4.5cm]{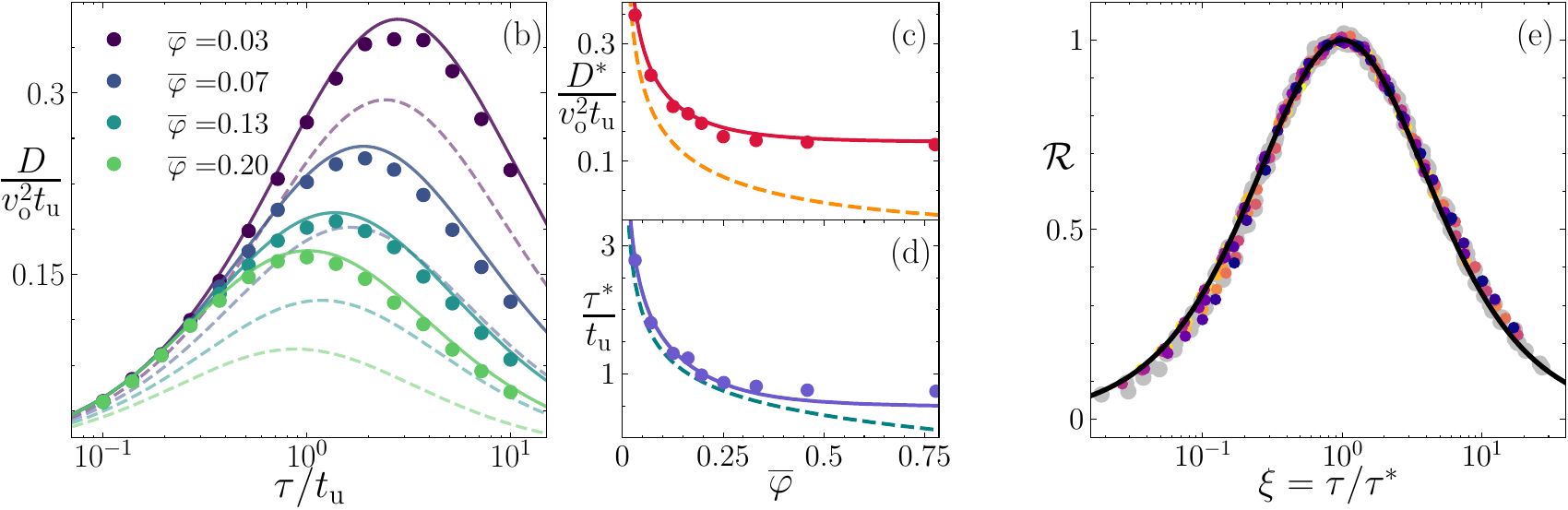}  
\vspace*{-2mm}

\caption{
{\bf Run-and-tumble with sliding along surface.} 
(a)~The porous medium consists of disks on a square lattice with spacing~$W$. 
The microorganism moves through porous space (mode 1) 
or along obstacle surfaces (modes~$2$ and~$3$). 
It aligns with walls upon collision and when tumbling at the surface, escapes with probability $1/2$ and reverses with probability $1/4$. 
(b)~Diffusivity as a function of mean run time, comparing numerical data (points) and model (solid line).  
Dashed lines show diffusivity without sliding. Time unit is $\Tunit = W /\vo$.  
(c-d)~Diffusity maximum~$\difs$ and optimal mean run time $\taurs$ as a function of solid fraction, 
with and without surface sliding (continuous and dashed lines respectively). 
Circles show simulation data. 
(e)~The modified diffusivity ratios $\rat$ for various solid fractions ($\porb=0.08-0.69$) and swimming strategies 
collapse on the master curve Eq.~\eqref{eq:rat} (black line). 
Grey circles show additional data for sliding without surface reversals, 
sliding on ordered rectangles and sliding in disordered environments.  
The 37 datasets are detailed in Table~II of \SM.  
}  
\label{fig:3}
\end{figure*}

 
\paragraph*{Multimodal motion and surface sliding.} 
In addition to trapping at the wall studied so far (and extended to partial trapping in \SM),  
a relevant class of surface behaviors involves sliding, 
wherein a microorganism encountering the solid moves with velocity $\vo$ along the surface.  
Motion then becomes multimodal. 
To understand the implications, 
we first investigate the diffusivity of a bimodal process where two modes $m=1$ and $2$ alternate. 
Mode~$m$ is characterized by a Liouvillian  $\liou_m$, 
which governs the Fokker-Planck equation $\partial_t p_m = \liou_m p_m$ for the distribution $p_m(\theta,t)$ 
of the angle $\theta$ describing the microorganism orientation (we took $\dm=2$ for simplicity).
In particular,  
motion that includes rotational diffusion, 
tumbles with rate $\taur^{-1}_m$ and a turning angle distribution~$h_m$ 
yields $\liou_m = \drm{m} \partial_{\theta \theta} - \taur^{-1}_m(1 - h_m \otimes)$, 
where $\otimes$ denotes convolution. 
A transition from mode $m$ to $m'$ occurs at rate~$\mu_m$ 
and induces a reorientation specified by the distribution of turning angle $h_m$ with correlation~$\be_m$.   

Using an exact formalism based on Fourier and Laplace transforms~\SM, 
the diffusivity of a bimodal process is:       
\begin{align}
\dif = \nu_1 D_1(\mu_1, \bee_1) + \nu_2 D_2(\mu_2, \bee_2) + (\be_1 + \be_2) C. 
\label{eq:bim}
\end{align}
Here, 
 $\nu_m=\taur_{m}/(\taur_{m}+\taur_{m'})$ 
is the fraction of time spent in mode~$m$. 
$D_m(\mu,\be) \eqdef \frac{1}{2} \suml [- \liou_m(l) +\mu_m \beb]^{-1}$, 
with $\liou_m(l)$ the Fourier series of $\liou_m(\theta)$,  
is the diffusivity for a process involving motion in mode~$m$, 
but  interrupted with rate~$\mu$ by reorientation events of correlation~$\be$. 
The effective correlation
$\bee_m \eqdef (\be_1 \be_2/2) \suml [1 - \liou_{m'}(l)/\mu_{m'}]^{-1}$  
characterizes the reorientation occurring  between an escape from mode~$m$ and a return to mode~$m$.   
The correction~$C$ is given by    
$ C^{-1} \eqdef  (\mu_1+\mu_2) (\left[ \mu_1 \mu_2 \difr_1(0) \difr_2(0)\right]^{-1} - \be_1 \be_2)$
where $\difr_m(0) \eqdef \difr_m(0,.)$ is the diffusivity in pure mode~$m$. 
The first two terms in Eq.~\eqref{eq:bim}
are the weighted averages of interrupted unimodal diffusivities,
because each mode~$m$ is interrupted with rate $\mu_m$ by the other, 
and these interruptions induce  a reorientation with effective \crl{}~$\bee_m$. 
Yet, Eq.~\eqref{eq:bim} also includes an additional correction term whenever some degree of correlation is retained, on average, when switching mode ($\be_1 + \be_2 \neq 0$).

As a specific case coupling bulk and surface motion, 
we consider a square lattice of disks  
and assume that tumbles at the surface can induce either escapes or reversals along the surface.  
Since sliding around obstacles can be clockwise or counter-clockwise, 
the process is now trimodal (Fig.~\ref{fig:3}a).  
The formalism 
can be extended~\SM{} and yields a formula analogous to Eq.~\eqref{eq:bim}. 
The  predicted diffusivity is in agreement with simulation data (Fig.~\ref{fig:3}b). 
At high solid fraction~$\porb$, 
the maximum $\difs$ vanishes in case of wall trapping, 
whereas for sliding it can display non-monotonic behavior or reach a plateau  (Fig.~\ref{fig:3}c), 
where displacement becomes surface-dominated.   
The optimal run times 
with sliding and trapping remain comparable up to 
$\porb \simeq 0.4$,  
which suggests that environments with distinct surface properties 
may nevertheless lead to similar optimal strategies.

%
Finally, we show that the universal law of dispersal 
is also relevant to multimodal motions.  
Because analytical approaches  become too complex,  
we resort to numerical simulations to explore several variations 
in surface behaviors and environments. 
These include  sliding on disks without surface reversals, sliding on rectangles with reversals at corners, 
as well as monodisperse ordered disks and several types of polydisperse disordered disks. 
Taking~$c$ as a free parameter in Eq.~\eqref{eq:rat}, 
all scenarios considered lead to a collapse on the master curve $\rat(\xi)$ (Fig.~\ref{fig:3}e).   
This  indicates that the law of dispersal,  demonstrated above for trapping at surfaces, 
also extends to  a variety of surface behaviors  that involve sliding.

%
In spite of its apparent broad applicability, 
the law of dispersal is not without exception. 
Finding a generic criterium for failure is a challenge,  
but one requirement is indicated by the following counter-example. 
Assume a microorganism that moves amid ordered rectangular obstacles, 
slides along surfaces and escapes at every corner with direction unchanged. 
Because motion is essentially ballistic  and tumbles are not any more needed for escape, 
dispersal is highest  for $\taur \to \infty$ 
and there is no maximum at finite run time.  
One necessary condition for collapse 
is thus the existence of a trade-off that penalizes both short and long run times. 

\paragraph*{Discussion.} 

Though applied to a RT microorganism, 
the model is  applicable to non-tumbling motion. 
For an active Brownian particle (ABP) with vanishing surface velocity 
and which escapes by rotational diffusion~\cite{Moen_prr-2022},  
Eq.~\eqref{eq:dif} for diffusivity still holds, 
with $\xi=\taup/\taups$ and  $\taup \eqdef 1/\dr'$ the persistence time~\SM.   
For all natural porous environnements where surface trapping is a good approximation, 
such as rocks, soils, gels and  tissues~\cite{Bhattacharjee_sm-2019,Bhattacharjee_natcom-2019}, all having irregular boundaries, 
the diffusivity given by Eqs.~\eqref{eq:dif}-\eqref{eq:rat} is  remarkably generic. 
To predict long-time dispersal, 
most properties of the microstructure~\cite{book_Torquato-RandomHeteroMat} are irrelevant:   
only the mean chord length matters.  

%
The connection between dispersal and Cauchy universality sheds new light 
on some previous findings. 
The idea of a generic dispersal in porous media 
was suggested by Mattingly~\cite{Mattingly_arxiv-2023} 
from the study of a specific system~\footnote{
Specifically, the system considered involves 
a single type of RT motion -- ballistic runs and isotropic tumbles --
within a specific porous medium  -- randomly placed overlapping disks.}.  
His prediction, 
recovered as a particular case of Eq.~\eqref{eq:model}, 
was obtained by homogeneization. 
Our derivation from a minimal model    
explains why the result is  widely applicable.  
Besides, 
for RT polymers in a disordered medium, 
Ref.~\cite{Kurzthaler_natcom-2021} concluded that the ``size of the pores, not their shape, matters'' 
while for  ABPs in a periodic medium,  
Ref.~\cite{Dehkharghani_comphys-2023} 
identified the ``effective mean free path as the critical length scale governing cell transport''. 
Both statements follow from the  Cauchy universality embodied in Eq.~\eqref{eq:lc}.    
Finally,  
the reverse-when-stuck strategy was shown numerically to outperform other swimming patterns~\cite{Lohrmann_pre-2023}. 
Equation~\eqref{eq:model} allows to generalize this conclusion.   
The optimal pattern 
for an organism with surface sensing ability 
involves ballistic runs and escape immediately after collision 
 in a direction parallel -- not normal -- to the surface~\footnote{  
Taking $\tauw=0$ in Eq.~(1) maximizes the first term,  
while the second is largest for  ballistic motion and $\beb$ small.}. 

%
%


To conclude, 
we found that the dispersal of motile microorganisms within porous media 
is governed by a generic law, 
whose universality originates in the invariance of the mean free path.  
Cauchy universality has been shown to  govern  wave propagation through scattering media~\cite{Pierrat_pnas-2014,Savo_sci-2017} 
and  residence time of bacteria in microstructures~\cite{Frangipane_natcom-2019}.   
It also  implies that, 
whatever the diversity of motion patterns and porous media,   
microbial dispersal can be understood within a unified framework.  
Future research may
assess the effect of non-Poissonian processes
for tumble and trapping~\cite{Detcheverry_epl-2015,Bhattacharjee_natcom-2019}, 
and characterize anisotropic dispersal induced by external fields, flows~\cite{deAnna_natp-2021} or  
symmetry-breaking microstructures~\cite{Potiguar_pre-2014}.   
Finally, 
given the analogy between random motions 
and polymer chains, 
one may wonder about the implications of Cauchy invariance 
for  polymers in porous media  and nano-composites~\cite{Kumar_jcp-2017,Jancar_polymer-2010,Zeng_pps-2008}.


%

%



\vspace*{1mm}

\paragraph*{Acknowledgments.}
We acknowledge financial support from ANR-20-CE30-0034 BACMAG and ETN-PHYMOT within the European Union’s Horizon 2020 research and innovation programme under the Marie Sk\l odowska-Curie grant agreement No 955910. 

\vspace*{3mm}

$^*$ Equal contribution. 

\bibliographystyle{bibgenym}
\bibliography{man-bib,man-bibspe}   

\end{document}